\documentclass[sn-mathphys,Numbered]{sn-jnl}

\usepackage{graphicx}%
\usepackage{multirow}%
\usepackage{amsmath,amssymb,amsfonts}%
\usepackage{amsthm}%
\usepackage{mathrsfs}%
\usepackage[title]{appendix}%
\usepackage{xcolor}%
\usepackage{textcomp}%
\usepackage{manyfoot}%
\usepackage{booktabs}%
\usepackage{algorithm}%
\usepackage{algorithmicx}%
\usepackage{algpseudocode}%
\usepackage{listings}%

\theoremstyle{thmstyleone}%
\theoremstyle{thmstyletwo}%

\theoremstyle{thmstylethree}%

\colorlet{darkgreen}{green!50!black}
\colorlet{brightyellow}{yellow!75!red}
\colorlet{orange}{red!50!yellow}
\colorlet{darkblue}{blue!60!black}
\colorlet{darkred}{red!80!black}

\usepackage[normalem]{ulem}

\begin{document}

\title[Article Title]{Abnormal solutions of Bethe--Salpeter equation with massless and massive exchanges}


\author*[1]{\fnm{Jaume} \sur{Carbonell}}\email{jaume.carbonell@ijclab.in2p3.fr}

\author[2]{\fnm{Vladimir A.} \sur{Karmanov}}\email{karmanovva@lebedev.ru}
\equalcont{These authors contributed equally to this work.}

\author[2]{\fnm{Ekaterina A.} \sur{Kupriyanova}}\email{kupr\_i\_k@mail.ru}
\equalcont{These authors contributed equally to this work.}

\author[1]{\fnm{Hagop} \sur{Sazdjian}}\email{hagop.sazdjian@ijclab.in2p3.fr}
\equalcont{These authors contributed equally to this work.}

\affil*[1]{\orgdiv{Universit\'e Paris-Saclay, CNRS/IN2P3, IJCLab, 91405 Orsay,France}}

\affil[2]{\orgdiv{Lebedev Physical Institute, Leninsky prospect 53, 119991 Moscow, Russia}}



\abstract{We  summarize the main properties of the so called ''abnormal solutions'' 
of the Wick--Cutkosky  model, i.e. two massive scalar particles interacting via  massless scalar exchange ("photons"), 
within the Bethe--Salpeter equation.
These solutions do not exist in the non-relativistic limit, in spite of having very small binding  energies. 
They present a genuine  many-body character dominated by photons, 
with a norm of the valence constituent wave function (two-body norm) that vanishes in the limit of zero binding energy.

We present  new results concerning the massive-exchange case, in particular
determine under which conditions is it possible to obtain such peculiar solutions without spoiling the model by tachyonic states ($M^2<0$).}





\maketitle


\section{Introduction}\label{Sec_Intro}

Lorentz invariance of a physical theory does not only manifest itself
when the velocities of the particles are comparable to the speed of light or when their momenta are comparable to the constituent rest masses.
This fundamental symmetry of nature can also have dynamical
consequences in the low-energy limit and can induce quantitative and qualitative differences with respect to a non-relativistic description.

One could expect, indeed, that  when describing a bound state of two particles
with a very small binding energy, and involving momenta smaller than the constituent masses, both 
approaches, a relativistic and a non-relativistic one, would lead to very similar results. 
However this is not always the case.

For instance, when considering the zero-binding-energy limit of the Light-Front \cite{CDKM_PREP300_1999}  and of the 
Bethe--Salpeter (BS) \cite{BS_51,SB_51}   equations
in a $\varphi^2\chi$ scalar theory, it is found \cite{MC_PLB474_2000}  that the results of these covariant theories are very close to each other,  but differ
from the results of the non-relativistic Schr\"odinger equation when the mass $\mu$ of the exchange particle is non-zero. 

More spectacular is the fact that there are families of low-energy solutions that
exist in a relativistic theory while they are totally absent in its non relativistic limit.
This happens within the BS equation when considering two  scalar particles interacting
by a massless ($\mu$=0) scalar exchange.

The properties of such states, first discovered by Wick \cite{Wick_PR96_1954} and Cutkosky \cite{Cutkosky_PR96_1954} in the $\mu$=0 case,
and since denoted "abnormal solutions", is the main subject of the present contribution,
with special emphasis on its eventual persistence in the massive ($\mu>0$) case.

We present in Section \ref{Sec_BSE} a brief  summary of the BS equation for the scalar model.
Section \ref{C_Solution} is aimed to describe the Cutkosky solution for the massless-exchange case and  the properties of normal and abnormal states.
Section \ref{Sec_mu_ne_0} contains  selected results of the massive-exchange case.
Concluding remarks follow in \ref{Sec_Conclusions}.

\section{The Bethe--Salpeter equation}\label{Sec_BSE}

The BS equation deals with a well-defined object form the  Quantum Filed Theory point of view: the matrix element of the T-product of the Heisenberg operators taken between the vacuum and the bound state \cite{GML_PR84_51} 
\begin{equation}\label{BSA}
 \Phi(x_1,x_2,P)= <0\mid T \{ \phi(x_1)\phi(x_2) \}  \mid P >  
 \end{equation}
 
Its Fourier transform $ \Phi(k,P) $
\[  \Phi(  x_1,x_2,P) =  \int {dp_1\over(2\pi)^4}  {dp_2\over(2\pi)^4}  \Phi(p_1,p_2) \; e^{-iPx}   \; e^{-ikx}   =  \; e^{-iPx}    \int {dk\over(2\pi)^4}   \Phi(k,P)    \; e^{-ikx}       \]
written in terms of the total    $P= p_1+p_2$ and relative $ 2k=p_1-p_2 $ momenta, obeys the equation
\begin{equation}\label{BS_Phi}
\Phi(k,P) = S_1(k,P)\; S_2(k,P) \;  \int{d^4k'\over(2\pi)^4} \; iK(k,k';P)\; \Phi(k',P)  
\end{equation}
where 
\begin{eqnarray*}
S_1(k,P)  &=&  \frac{i} {\left({P\over2}+k\right)^2-m^2+i\epsilon}   \cr
S_2(k,P)  &=&  \frac{i} {\left({P\over2}-k\right)^2-m^2+i\epsilon}  
\end{eqnarray*}
 are the free particle propagators in the case of two equal masses and $iK$ is the interaction kernel. 
 If $iK$  contained all irreducible graphs of a Lagrangian density, the solution of \eqref{BS_Phi} would be equivalent
 to the solution of the full QFT problem. This kind of "mantra" is however a wishful thinking, not only because
 nobody knows how to construct such a kernel, but, would it be the case, the corresponding integral equation would not be integrable. 
One is then limited to use very simple reductions that keep only a vague flavour of the underlying Lagrangian theory. 
In the simplest case of two scalar particles  of mass $m$ interacting via  a massive scalar exchange $\mu$  the ladder kernel reads
 \begin{equation}\label{Kernel}
  iK(k,k')= - { g^2\over (k-k')^2  - \mu^2 + i \epsilon}    \quad \Longrightarrow \quad V(r)=-{g^2\over4\pi}  \; {e^{-\mu r} \over r}
  \end{equation}
 In the non-relativistic limit it leads to the Yukawa potential $V(r)$. 

Equation \eqref{BS_Phi} is an implicit eigenvalue equation with repect to the total mass squared of the system, $M^2\equiv P^2$, which in this ladder approximation,
appears only in the free propagators.

\begin{figure}[htbp]
\centering
\includegraphics[width=5.cm]{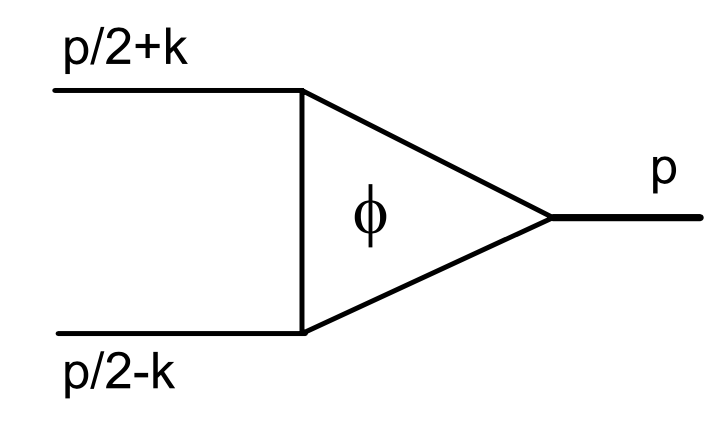}
\caption{Feyman graph representing the BS amplitude in momentum space.}\label{Fig_Feynman}
\end{figure}

The BS amplitude $\Phi(k,P)$ is quite a nasty mathematical object, plagued with singularities, as a function of the arguments, 
which represents the Feynman amplitude depicted in Fig. \ref{Fig_Feynman} and  which has not an easy interpretation in terms of wave functions.
These singularities motivated G. Wick \cite{Wick_PR96_1954}	 to change the  original formulation in Minkowski space into the Euclidean metric,
introducing the, since then famous, Wick rotation in the time-like component: $k_0$=$ik_4$ such that  
$k_M=(k_0,k_1,k_2,k_3)\to k_E=(k_1,k_2,k_3,k_4=-ik_0)$ 
and $k_M^2$=$k_0^2-k_1^2-k_2^2-k_3^2$=$-k_E^2$=$-(k_1^2+k_2^2+k_3^2+k_4^2)$. 
This change of metric paved the way for obtaining the first solutions of BS equation with kernel \eqref{Kernel} with $\mu$=0.
Results  were published by Wick himself in the quoted reference \cite{Wick_PR96_1954} and 
in a much more complete way by Cutkosky  in the subsequent paper of the same journal \cite{Cutkosky_PR96_1954}.
They constitute the so-called {\bf Wick--Cutkosky (W-C) model, although we will abusively use the same denomination} for
the $\mu\ne$0 case.

For the S-wave and the W-C model, the equation reads:
\begin{small}
\begin{equation}\label{EBSE}
 \left[ ( k^2+ k^2_4 +m^2-\frac{1}{4}M^2)^2+M^2k^2_4 \right]  \Phi_E(k_4,k) =
 \int_0^{\infty} dk'\int_{-\infty}^{\infty} dk'_4 V^E_0(k_4,k;k'_4,k')    \Phi_E(k'_4,k')  
\end{equation}   
\end{small}
where the S-wave Euclidean kernel 
\begin{eqnarray}\label{Vt0c}
V^E_0(k_4,k;k'_4,k')  = \frac{\alpha m^2 k'}{\pi^2 k}  \;  \log\frac{[(k_4-k'_4)^2+(k+k')^2+\mu^2]}{[(k_4-k'_4)^2+(k-k')^2+\mu^2]}.
\end{eqnarray}
is  smooth everywhere for $\mu\ne0$ and presents only  logarithmic singularities in the diagonal $\{k'=k\} \times \{k'_4=k_4\}$ for $\mu=0$. Here, $\alpha=g^2/(16\pi m^2)$.
Nowadays there exist several methods to solve accurately  the BS equation in Euclidean space for a bound state problem.
They apply to a large variety of kernels with bosons and fermions and even beyond the ladder approximation.
A much more precarious situation is however observed in the scattering problem.

Other methods have been developed aiming to obtain a Minkowski  solution of the same equation. They are based
on an integral representation of the BS amplitude which collects all the singularities
in an analytic term and deals with a regular weight function that obeys a modified BS equation.
They are  also much better adapted to understand the abnormal states for the massless as well as for the massive exchange cases. 
These "Minkowski-space" methods are widely inspired by the one developed by Cutkosky  in his solution of the massless W-C model,
that will be briefly  summarized in the coming section.

\section{Cutkosky's solution for the massless case}\label{C_Solution}

\bigskip
In his first publication \cite{Cutkosky_PR96_1954}, Cutkosky  searched for the solution of \eqref{BS_Phi}  with the interaction
kernel \eqref{Kernel} in the form of an integral representation:
\begin{equation}\label{Sol_WC}
\Phi^{LM}_{n}(k,P)=  \sum_{\nu=0}^{n-L-1}  \; \int_{-1}^{+1} dz \; g_{nL}^{\nu}(z) \;  \frac{Y_{LM}(\hat k) }{ \left[  k^2 + (k\cdot P)\;z - \left(m^2 - {P^2\over 4} \right) + i \epsilon \right]^{2+n-\nu}  }
\end{equation}
in terms of  unknown ``weight functions'' $g_n^{\nu}(z)$.

For the S-wave, which is the only case we are going to consider here, and disregarding a global normalization factor, (\ref{Sol_WC}) turns into
\begin{equation}\label{Sol_WC_S}
\Phi_{n}(k,P)=   \sum_{\nu=0}^{n-1}  \; \int_{-1}^{+1} dz \;  \frac{g_n^{\nu}(z)}{ \left[  k^2 + (k\cdot P)\;z - \left(m^2 - {P^2\over 4} \right) + i \epsilon \right]^{2+n-\nu}  }
\end{equation}

The solution $\Phi_n$, where $n=1,2,3,\ldots$ , which plays the role of the principal quantum number
in the non-relativistic Coulomb problem, is obtained as a superposition of $n$ components  $g_n^{\nu}(z)$ labelled by $\nu=0,1,\ldots,n-1$.

Cutkosky  obtained  a coupled set of integral equations for the weigh functions $g_{n}^{\nu}(z)$  in the  -- highly inspired but useless -- form\footnote{Note that there was a misprint in the integration limits over t in the original publication \cite{Cutkosky_PR96_1954}.}
\begin{eqnarray}
g_n^{\nu}(z) &=& {\lambda\over2}  \sum_{\nu'=0}^{\nu} \frac{ (n-\nu+1)! \; (n-\nu'-1)! }{ (n-\nu'+1)! \; (n-\nu-1)!}   \;   \int_{-1}^{+1} dt \; \int_0^1 dx \;  x(1-x)^{n-\nu-1}  \cr
 &\times &\; \int_{-1}^{+1}dz'  \frac{\delta[ z -xt-(1-x)z'] }{ (1-\eta^2 + \eta^2 z'^2  )^{\nu-\nu'+1}}  \; \; g_n^{\nu'}(z')   \label{Cut_Sol_L0}
\end{eqnarray}
where
\begin{eqnarray}
\lambda&=& {g^2\over 16\pi^2 m^2}   =  {\alpha\over \pi}   \label{lambda} \\
\eta       &=& {M\over 2m} = 1- {B\over 2m}  \label{eta}
\end{eqnarray}
After simplifying the coefficients and introducing
\begin{equation}\label{Q}
 Q(z)= 1-\eta^2(1-z^2)
 \end{equation}
one is left with the, still useless,  system of integral equations
\begin{eqnarray}\label{Cut_Sol_L0_2}
g_n^k(z) &=& {\lambda\over2}  \sum_{k'=0}^k  \frac{ (n-k+1)(n-k)} {  (n-k'+1) (n-k')}  \cr
&\times& \int_{-1}^{+1} dt\int_0^1 dx \;  x(1-x)^{n-k-1} \; \int_{-1}^{+1}dz'  \frac{\delta[ z -xt-(1-x)z'] }{ Q(z')^{k-k'+1}} \;  g_n^{k'}(z') 
\end{eqnarray}
The integration over $dt$ can be performed, using the $\delta$ function 
\begin{enumerate}
\item Since $x>0$
\[ \int_{-1}^{+1} dt \; \delta[ z -xt-(1-x)z'] = \int_{-1}^{+1} dt \; \delta[xt -  (z -(1-x)z') ] = {1\over x}   \int_{-1}^{+1} dt \; \delta\left[t -  {z -(1-x)z' \over x}  \right]  \]
which takes the form
\begin{equation}\label{delta}
\int_{-1}^{+1} dt \; \delta(t - t_0 )  \qquad  t_0(x,z,z')=  {z -(1-x)z' \over x}     
\end{equation}
\item Expression (\ref{delta}) vanishes except if  $t_0\in [-1,+1]$ that is if the following two conditions are satisfied
\[ -1 < \frac{z-(1-x)z'}{x}  < +1 \]
rewritten as
\begin{equation}\label{condition}
 -1 < {z-z'\over x}  + z'< +1 
 \end{equation}
To fulfill these conditions we must distinguish the sign of $z-z'$
\begin{itemize}
\item If $z-z'>0$,\par
- the left part of (\ref{condition}) is automatically fulfilled 
\footnote{since ${z-z'\over x} >z-z'$ and so  ${z-z'\over x}+z' >z>-1$}\par
- the right part requires the condition
\[ x> { z-z' \over 1-z'}\]
and so
\begin{eqnarray}
 \int dt \; \delta(t - t_0) \int_0^1 dx\; x(1-x)^{n-k-1} &=& \int_{ { z-z' \over 1-z'}  }^1 dx \; (1-x)^{n-k-1} \cr
 &=&  -  { 1\over n-k} \left[  (1-x)^{n-k}\right]_{  { z-z' \over 1-z'}}^1  \cr
 &=&  { 1\over n-k} \left(  {1-z \over 1-z'} \right)^{n-k}   \label{zmzp_gt0}
\end{eqnarray} 
\item If $z-z'<0$,\par

- the right part of  (\ref{condition}) is automatically fulfilled since $ z'+ {z-z'\over x}<z'<+1$\par
- the left part requires 
\[ x> {z'-z\over 1+ z'} \]
\begin{eqnarray}
 \int dt \delta( t - t_0  ) \int_0^1 dx \; x(1-x)^{n-k-1}   &=& \int_{ { z'-z \over 1+z'}  }^1 dx \; (1-x)^{n-k-1} \cr
 &=&  -  { 1\over n-k} \left[  (1-x)^{n-k}\right]_{  { z'-z \over 1+z'}}^1  \cr
 &=&  { 1\over n-k} \left(  {1+z \over 1+z'} \right)^{n-k}   \label{zmzp_lt0}
\end{eqnarray}  
\end{itemize}

\end{enumerate}
Gathering (\ref{zmzp_gt0}) and (\ref{zmzp_lt0}) we obtain
\[ \int dt \delta( t - t_0  ) \int_0^1 dx \; x\; (1-x)^{n-k-1}= {1\over n-k} R^{n-k}(z,z')    \]
where we have introduced  the kernel
\begin{equation}\label{R}
R(z,z')= \theta( z-z')  \left( {1-z \over 1-z'} \right) + \theta( z'-z)\left({1+z \over1+z'} \right)= \left\{  \begin{array}{lcl}  {1-z \over  1-z'},\quad &\mbox{if}&\quad z'<z  \cr   {1+z \over  1+z'},\quad &\mbox{if}&\quad z'>z \end{array}  \right.
\end{equation}

\begin{figure}[h!]
\begin{center}
\centering\includegraphics[width=6.cm]{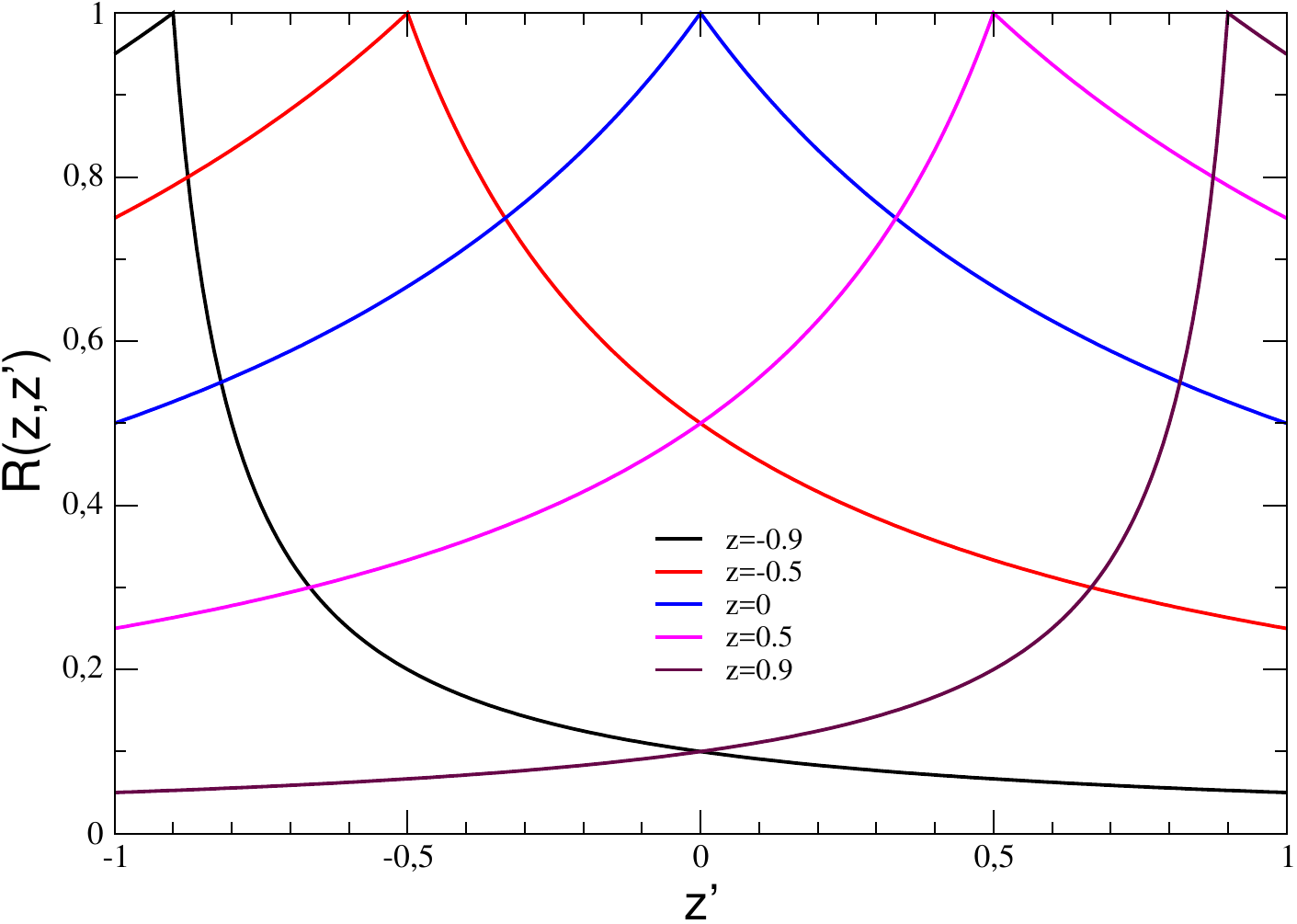}\hspace{0.2cm}
\centering\includegraphics[width=6.4cm]{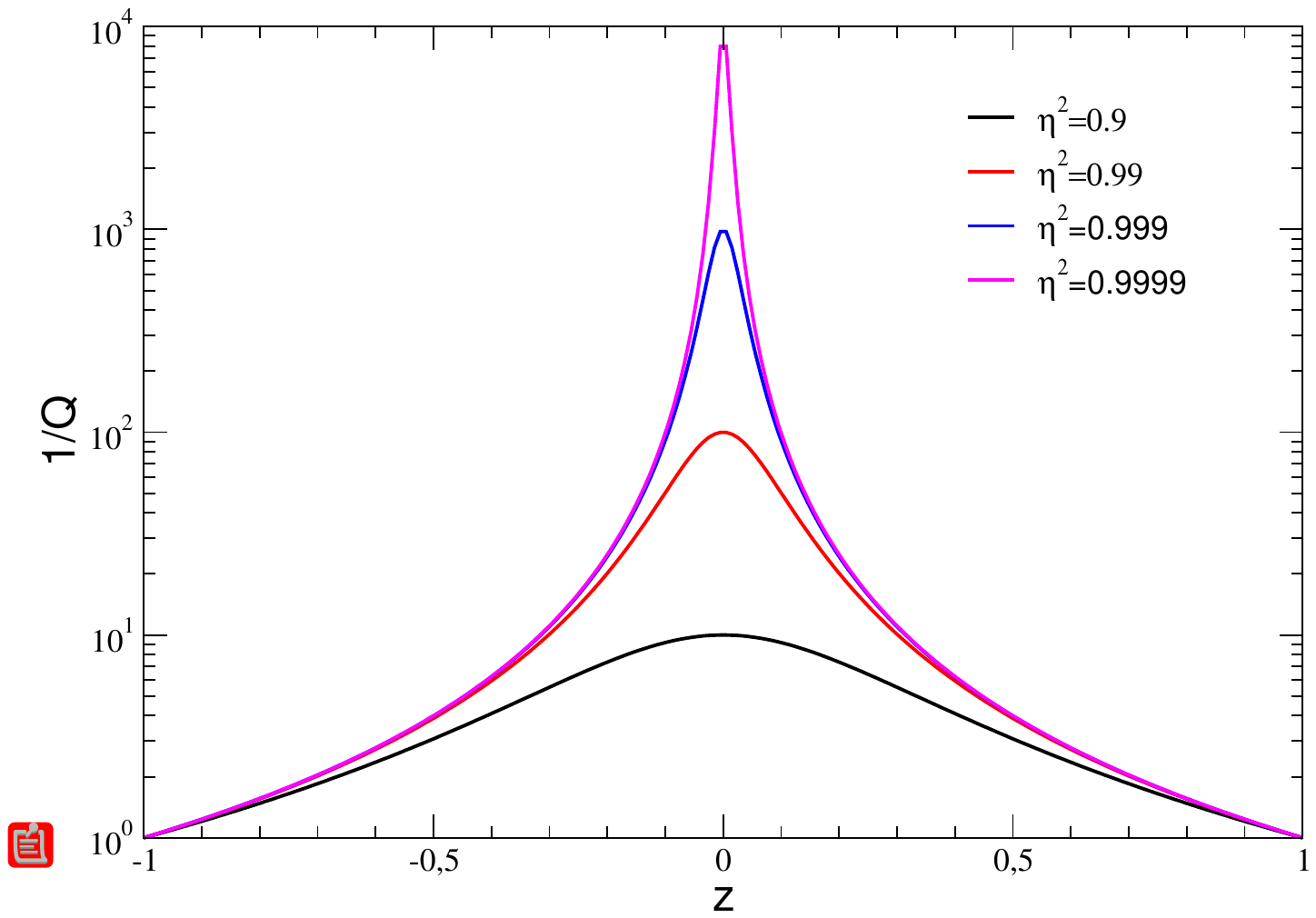}
\caption{Left panel  kernel R(z,z')  defined in \eqref{R} and right panel kernel $Q$ defined in \eqref{Q}  for different values of $\eta$.}\label{Fig_R_Q}
\end{center}
\end{figure}

\noindent
By inserting this expression in (\ref{Cut_Sol_L0_2}), one is finally left with the following triangular system of coupled one-dimensional integral equations,
useful for numerical calculations:  
\begin{equation}\label{CEQ_L0}
 g_n^{\nu}(z)=  {\lambda\over2}  \sum_{\nu'=0}^{\nu}   c_n^{\nu\nu'} \; \int_{-1}^{+1} dz' \; {  R^{n-\nu}(z,z') \over Q^{\nu-\nu'+1}(z') } \; g_n^{\nu'} (z')  \qquad \nu=0,1,...n-1 
\end{equation}
with coefficients
\[  c_n^{\nu\nu'}=\frac{ (n-\nu+1)} {  (n-\nu'+1) (n-\nu')}  \]
 The kernels $R(z,z')$ and $Q(z)$, defined respectively in \eqref{R} and  \eqref{Q}, are displayed in Fig. \ref{Fig_R_Q}.
$R(z,z')$ is continuous in both arguments  but has a  cusp at $z'=z$, while $1/Q(z)$ is peaked around $z=0$. 
This peak becomes increasingly sharp when the binding energy tends to zero.
Since both kernels appear in  equation \eqref{CEQ_L0} to some integer power, when this power is large they become quasi-singular around
the critical points ($z'=z$ and $z=0$), and make the calculations difficult.

\noindent
For $n=1$ there is a single equation determining the unique component of $\Phi_1$
\begin{equation}
g_1(z)=  {\lambda\over2}    \; \int_{-1}^{+1} dz' \; {  R(z,z') \over Q(z') } \; g_1  (z') 
\end{equation}
For $n=2$, the solution $\Phi_2$ is determined by two  components, $g_2^0$ and $g_2^1$ , satisfying
\begin{eqnarray*}
g_2^0(z)  &=&  {\lambda\over2}    \;\;\;{1\over 2} \int_{-1}^{+1} dz' \; {  R^2(z,z') \over Q(z') } \; g_2^0  (z')   \cr
g_2^1(z)  &=&  {\lambda\over2} \left[    {1\over3}  \; \int_{-1}^{+1} dz' \; {  R(z,z') \over Q^2(z') } \; g_2^0  (z')  +    \int_{-1}^{+1} dz' \; {  R(z,z') \over Q(z') } \; g_2^1  (z')  \right]
\end{eqnarray*}
For $n=3$ the solution is given by 
\begin{small}
\begin{eqnarray*}
g_3^0(z) &=& {\lambda\over2}    \;\;\; {1\over 3} \int_{-1}^{+1} dz' \; {  R^3(z,z') \over Q(z') } \; g_3^0  (z')   \cr
g_3^1(z) &=& {\lambda\over2} \left[ c_3^{10} \; \int_{-1}^{+1} dz' \; {  R^2(z,z') \over Q^2(z') } \; g_3^0  (z')  +  c_3^{11} \int_{-1}^{+1} dz' \; {  R^2(z,z') \over Q(z') } \; g_3^1  (z')      \right] \cr
g_3^2(z) &=& {\lambda\over2} \left[ c_3^{20} \; \int_{-1}^{+1} dz' \; {  R(z,z') \over Q^3(z') } \; g_3^0  (z')      +  c_3^{21}\int_{-1}^{+1} dz' \; {  R(z,z') \over Q^2(z') } \; g_3^1  (z')   + c_3^{22}  \int_{-1}^{+1} dz' \; {  R^2(z,z') \over Q(z') } \; g_3^2  (z')\right]
\end{eqnarray*}
\end{small}
 
It is worth noticing that, for each $n$, the component $g_{n}^{0}$ decouples from the rest. It is  determined by a single equation 
\begin{equation}\label{CEQ_L0_gn0}
 g_n^0(z)=  {\lambda\over2n}   \; \int_{-1}^{+1} dz' \; {  R^{n}(z,z') \over Q(z') } \; g_n^{0}(z') 
\end{equation}
and provides  the full  spectrum of the W-C model. The rest of the components are needed to reconstruct the full BS amplitude  and other observables, like, e.g., form factors.

Since \eqref{CEQ_L0_gn0} is homogeneous, the norm of  $g_n^0$ is not fixed by \eqref{CEQ_L0_gn0}, and can be only determined
by normalizing $\Phi_n$ in \eqref{Sol_WC_S}.
On the contrary, there is no choice in the norm of the components $g_n^{\nu>0}$, since
they obey an inhomogeneous equation with a source term proportional to $g_n^0$.

There is an equivalent formulation of  \eqref{CEQ_L0} in  differential form.
For instance, eq. \eqref{CEQ_L0_gn0} providing $g_n^0$ for any $n$, is equivalent to the differential equation
\begin{equation}\label{g_n}
{g''}_n^0(z) + {2(n-1)z\over 1-z^2} \; {g'}_n^0(z)  - {n(n-1)\over 1-z^2} \; g_n^0(z) +   {\lambda  \over (1-z^2)Q(z,\eta)} \;  g_n^0 (z)  = 0   
\end{equation}
with the boundary conditions $g_n^0(\pm1)=0$.

For n=2, and after having obtained $g_2^0$ by \eqref{g_n}, the component $g_2^1$ is  a solution of the inhomogeneous equation
\begin{equation}
 {g''}_2^1 + {\alpha\over\pi} {1\over (1-z^2) Q(z,\eta) } \; g_2^1 =  -{\alpha\over3\pi} \; {1\over (1-z^2)  Q^2(z,\eta)}  \; g_2^0  
\end{equation}
with $g_2^0$ as a source term, and so forth for the other $n$'s.

\subsection{Normal and abnormal states}\label{WC_Normal}

\bigskip
The W-C model has an infinite family of solutions ($\Phi_n,M_n$), which, for small values of the coupling constant $\alpha$, are "logarithmically tangent"
to the non-relativistic Coulomb bound-state spectrum  \cite{FFT}:
\begin{equation}\label{Bn_N}
B_n (\alpha)= {m\alpha^2\over 4n^2} \; \left[ 1 + {4\over\pi} \alpha\ln\alpha + o(\alpha^2) \right]
\end{equation}
where $M_n=2m-B_n$.

For a given $n$, the function $g_n^0(z)$ satisfies the homogeneous equation (\ref{g_n}), in which $n$ plays role of a parameter. Such equations usually have a whole spectrum of solutions. This just takes place in the equation (\ref{g_n}). Therefore, for each value of $n$=1,2,\ldots, there exists an additional series of eigenvalues labeled by a new quantum number $\kappa=0,1,\ldots$,
which is a consequence of the symmetry of the 4D Coulomb problem.
The components of the corresponding eigenstates  $g^{\nu}_{n\kappa}$ have a well defined parity determined by $\kappa$.
\[ g_{n\kappa}^{\nu} (-z)=(-)^{\kappa} g^{\nu}_{n\kappa}(z) \]

The subset $\kappa=0$ corresponds, for small values of $\alpha$,  to the standard Balmer series \eqref{Bn_N}, that is, to the "normal" non-relativistic solutions.
The other states, with $\kappa\ne0$ , do not have a counterpart in the non-relativistic theory and  for this reason were named by  Wick "abnormal states".
It is worth mentioning that the states with odd values of $\kappa$ have vanishing contributions to the S-matrix \cite{CM_PR140_1965}.
So the first abnormal state with dynamical content is $\kappa=2$.

\begin{figure}[h!]
\begin{center}
\centering\includegraphics[width=14.cm]{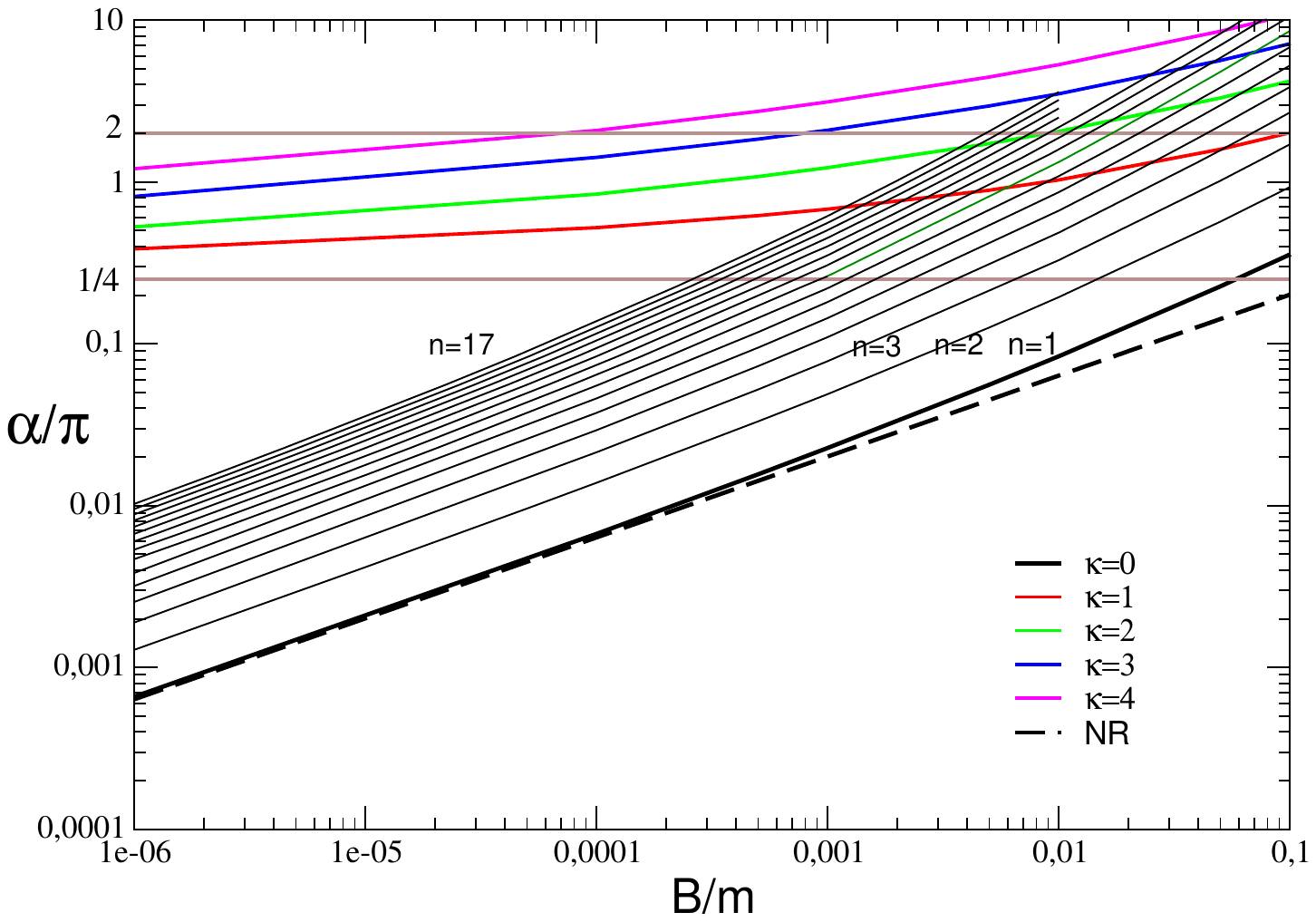} 
\caption{Solid black lines correspond to binding energies of the first (n=1,2,...) normal states ($\kappa=0$) 
as a function of the coupling constant $\alpha$. 
Abnormals states, limited to ground n=1, are in colored lines: $\kappa=1$ in red, $\kappa=2$ in green,... .}\label{Fig_lambda_B}
\end{center}
\end{figure}

We have displayed in Fig. \ref{Fig_lambda_B} the energies of the lower states as a function of the coupling constant $\lambda={\alpha\over\pi}$
in a log-log scale.
One sees clearly  that {\bf two different families of states} arise:  in solid black lines is the ensemble
of {\bf normal states} ($\kappa=0$) with different values of the principal quantum number $n$.
They have an accumulation point in ($\lambda,B$)=(0,0). We have included in black dashed line
the non-relativistic limit for the ground state.
Colored lines correspond to the ground state ($n=1$) of the {\bf abnormal solutions}: $\kappa=1$ in red, $\kappa=2$ in green,....
This family is totally decoupled, in its $\lambda(B)$ trajectory, from the first one, though the normal and abnormal series intersect. As seen in Fig. \ref{Fig_lambda_B}, between two excited normal states there are the abnormal ones. The abnormal series has an accumulation point at ($\lambda,B$)=(1/4,0). One
can  show \cite{Wick_PR96_1954,Cutkosky_PR96_1954} that,  for small values of B, the energies of the abnormal states behave as
\begin{equation}\label{Bn_A}
\lambda(B)\approx  {1\over 4}  + \frac{4\pi^2(\kappa-1)^2 } {\ln^2 {B\over m} }
\end{equation}
independently of the main quantum number $n$.
Notice from  Fig.  \ref{Fig_lambda_B} that the asymptotic value $\lambda_{min}=1/4$,  independent of  $\kappa$, is reached
very slowly when $B$ tends to zero. For instance, for $\kappa$=2 and B=10$^{-6}$, the value of $\lambda$ is still a factor two larger than $1/4$.
 
 The fact that this horizontal asymptote is reached always from above means that the abnormal solutions
exist only for large values of the coupling constant: $\lambda\ge {1\over 4}$ or $\alpha\ge{\pi\over 4}$.
This is the meaning of {\bf the horizontal bold brown line} at $\lambda=\lambda_{min}=1/4$.  

{\bf The existence of a minimal coupling constant to bind a two-body system is typical of a Yukawa-like massive-exchange potential}.
From this point of view,
the ground abnormal state ($\kappa=2$) behaves as if it was created by the exchange of a "massive photon" with 
a non zero effective mass $\mu_{eff}$. 
One can get an evaluation of its value by  means of the solution of the non-relativistic Yukawa model \cite{CdSK_Yukawa_FBS54_2013}.
One can see there that the minimal coupling constant for the appearance of the first bound state in the dimensionless problem  is  $G_0\approx 1.680$.
It is related to the minimal coupling constant of the dimensionful problem by $\alpha_{min}={\mu\over m} G_0$.
For all abnormal solutions one has $\alpha_{min}=\pi/4$, which gives  $\mu_{eff}/m=0.47$.

 On another hand, it turns out that for values of $\lambda>2$ the ground state of the model, i.e., the normal state with $n$=1 and $\kappa$=0, has $M^2<0$.
 If we want to restrict ourselves to non-tachyonic states, the coupling constant must be limited to $\lambda\le\lambda_{max}=2$. 
 This is the reason for the {\bf horizontal brown line} at $\lambda=\lambda_{max}$ in Fig.  \ref{Fig_lambda_B}.
 Since the first abnormal state is $\kappa$=2 and it crosses the line  $\lambda= \lambda_{max}$  at $B\approx 0.00903$, this means
 that all abnormal states, although requiring large values of the coupling constants to exist, concern very low-energy solutions.

It is  quite a paradoxical situation for a relativistic theory to predict a series of new low-energy states.
We would like to mention here that the existence of such states, would they be obtained with a simplified kernel, require the full covariance of the theory.
If the excitation in the time-like degrees of freedom are frozen, the abnormal states disappear \cite{Sazdzjan_JPG22_1996}.

\bigskip 
\subsection{Characterization of the abnormal states}\label{WC_Abnormal}

\bigskip
In a series of papers 
{\cite{KCS_LC2019,KCS_EPJWC204_2019,KCS_NTSE_2018}}, summarized in \cite{Abnormal_1_2021}, we have studied the properties
of the abnormal states in the W-C model, with the main aim of giving an intrinsic characterization of such states, rather than a relative position in the full spectrum of the model.

We have extensively compared the elastic and inelastic form factors of normal versus abnormal states.
The most striking difference  lies in the two-body contents of the corresponding state vectors.
Indeed, in the Light-Front approach, the state vector $\mid P\rangle$ appearing in the definition of
the BS amplitude \eqref{BSA} is a QFT state involving many-body components (Fock expansion):
\[ \mid P\rangle= \sum_{n\ge2}^{\infty} \Psi_n(k_1,k_2,\ldots,k_,) \mid n\rangle  \qquad \mid n\rangle= a^{\dagger}_{k_1} a^{\dagger}_{k_2} ,\ldots,a^{\dagger}_{k_{n-2}} b^{\dagger}_{q_1}  b^{\dagger}_{q_2} \]  
where the operators $b^{\dagger}$ and $a^{\dagger}$  are the creation operators of the constituent massive particles and of the exchanged 
particle, respectively, and where  $\Psi_n$  is, by definition, the n-body wave function.

The total norm of a state vector results from adding the partial norms of the corresponding  2-, 3- and many-body components
\[   \langle P\mid P\rangle= 1= \int \Psi_2^2  + \int \Psi_3^2 + \int \Psi_4^2 + \ldots = N_2+ N_3+ N_4 + \ldots \]
  
Having normalized the BS amplitude $\Phi$ by the condition $F$(0)=1, where $F(Q)$ is the elastic electric
form factor of the full bound system, we have obtained the two-body wave function by
projecting on the Light-Front plane, $\omega\cdot x$=0 with  $\omega^2=0$,  the BS amplitude
\[ \psi_2(k_1,k_2,P,\omega)= \frac{ (\omega\cdot k_1)  (\omega\cdot k_2)  }{\pi (\omega\cdot P) } \int_{-\infty}^{+\infty}  \Phi(k+\beta \omega,P) \; d\beta  \]
and by that, the two-body norm
\[ N_2 = \frac{1}{(2\pi)^3 } \int  \psi_2^2(k_{\perp},x) \frac{d^2 k_{\perp}dx}{2x(1-x)}.   \]

\begin{figure}[h!]
\begin{center}
\centering\includegraphics[width=6.3cm]{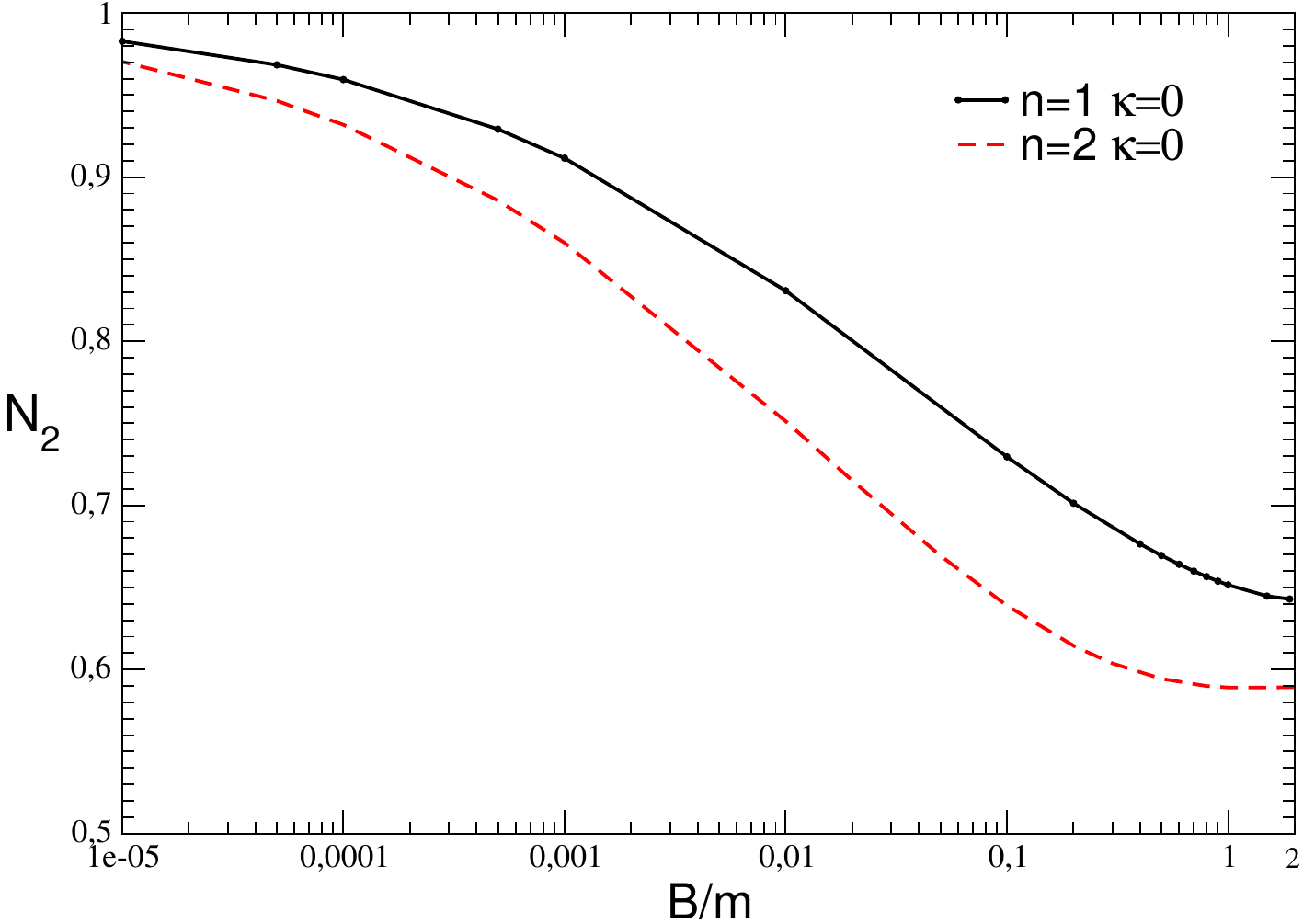}\hspace{0.2cm}
\centering\includegraphics[width=6.3cm]{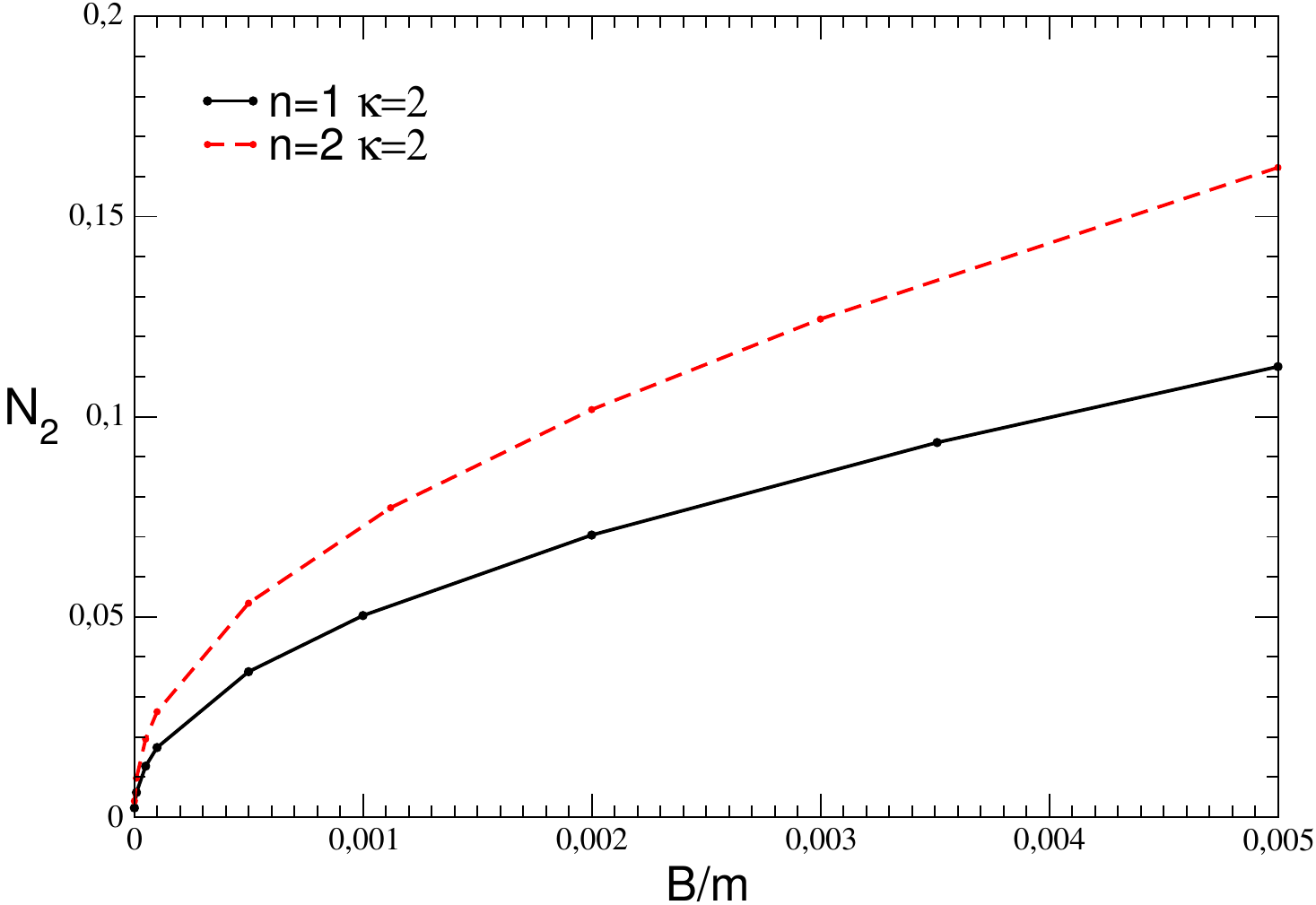}
\caption{$N_2$-dependence on the binding energy $B/m$ for the ground ($n=1$) and first excited ($n=2$) states. 
On left panel for the normal states ($\kappa=0$) and on right panel for the abnormal ones ($\kappa=2$).}\label{Fig_N2}
\end{center}
\end{figure}

We have displayed in Fig. \ref{Fig_N2} the two-body norm $N_2$ of the first two normal (left panel) and abnormal (right panel) 
states as functions of their binding energy $B$.
As one can see, an essential difference appears between the two-body contents of a normal and  an abnormal state. 
In the limit of small binding energies, the two-body norm $N_2$ of a normal state  tends to 1 (left panel) , indicating that  it is described by a 2-body.
valence wave function.
On the contrary, the abnormal states has a two-body norm that, in this $B\to0$ limit, tends to 0, indicating
a genuine many body character of such states.
This difference holds for the ground as well as for the first excited states.

This is an intrinsic difference between such kind of states, related to only their very internal structure and independent of their energy,
and was probably the most striking result of our previous work \cite{Abnormal_1_2021}.

It is worth mentioning here some kind of skepticism expressed by  Wick himself,
when commenting on the existence of such states at the very end of his seminal article \cite{Wick_PR96_1954}.
He wrote "About the possible existence of these abnormal solutions, we shall not try to speculate. 
Since they occur only for finite values of $\lambda$ ($\lambda\ge1/4$) it will be unwise to assume that they
are a property of the complete BS equation. Certainly the ladder approximation cannot be trusted to such extent".
To our knowledge, this judicious remark still remains an open question.
However the many-body character of the abnormal states that we have put in evidence, seems to provide an argument  in favour of their existence. It is hard
to imagine a possible mechanism with which the non-ladder many-body contributions could inhibit the construction of such  collective many-body states.

So far we have considered the case of equal constituent masses. 
If the Coulomb field would be provided by a heavy ion, interacting with an electron, 
we would deal with strongly unequal constituent masses. 
The existence of abnormal states in this case was studied in \cite{Cutkosky_PR96_1954,VAK_Un_m_EPJC84_2024}. 
It turns out that in a system with such different masses 
the abnormal states still exist. Moreover, the effect of unequal masses 
is attractive. 
The balance between the exchanged photons and the massive constituents is little sensitive to the mass ratio, 
and so the many-photon component still predominates.

It remains to see whether the abnormal states survive when the exchange mass differs from zero.
This will be the content of our forthcoming publication \cite{CKKS_EPJA2}	.
The first results are summarized in the next section.
 
\section{Solutions for the massive-exchange case}\label{Sec_mu_ne_0}

The solution of the massive-exchange case can be obtained by directly solving the BS equation
in Euclidean space \eqref{EBSE}.
There are however other alternatives which are inspired by the Cutkosky solution of the massless-exchange case, previously discussed. 
They are also based on a, now two-dimensional,  integral representation of the BS amplitude that is due to  Nakanishi \cite{Nakanishi} and that reads
 \begin{eqnarray}\label{bsint}
\Phi(k,p)&=&- i\int_{-1}^1dz' \int_0^{\infty}d\gamma   \frac{g(\gamma,z)}{\left[\gamma+m^2- \frac{1}{4}M^2-k^2-p\cdot k\; z-i\epsilon\right]^3}.
\end{eqnarray}
Once inserted in the BS equation, one can obtain an integral equation for the bi-dimensional spectral function $g(\gamma,z)$.
This approach was first suggested by Kusaka and Williams \cite{KW_PRD51_1995,KSW_PRD56_1997} 
with the aim of  obtaining  the BS solutions in Minkowski space,
and has  taken different forms during the almost thirty years of sustained developments 
\cite{bs1,bs2,FSV_PRD85_2012,FSV_PRD89_2014,FSV_EPJC75_2015,dPaPRD16,GutPLB16,CFK_PLB769_2017,Jia_PRD109_2024}.

Our last formulation  \cite{CFK_PLB769_2017}	 is based on a combined use of the Nakanishi representation, the Light-Front projection of the BS amplitude and the Stieltjes transform.
The (bound state) BS equation for the weight function $g$  takes the standard form

\begin{equation}\label{gNg}
g(\gamma,z)=\int_0^{\infty}d\gamma'\int_{-1}^{1}dz'\; N(\gamma,z;\gamma',z') \; g(\gamma',z')
\end{equation}
Several, a priori equivalent,  forms of  the kernel $N$ corresponding to the W-C model  can be found in 
\cite{FSV_PRD89_2014,CFK_PLB769_2017,VAK_Kernel_PRD104_2021}. 

Although the spectral function $g(\gamma,z)$ is smooth, the kernel $N$ has several moving singularities in both integration variables $\gamma',z'$.
They are of course integrable but must be treated  with some care to avoid  spurious structures. 
The detailed analysis of these singularities depends on the particular form of the kernel and will be given in  \cite{CKKS_EPJA2}.

\begin{figure}[htbp]
\begin{center}
\includegraphics[width=6.cm]{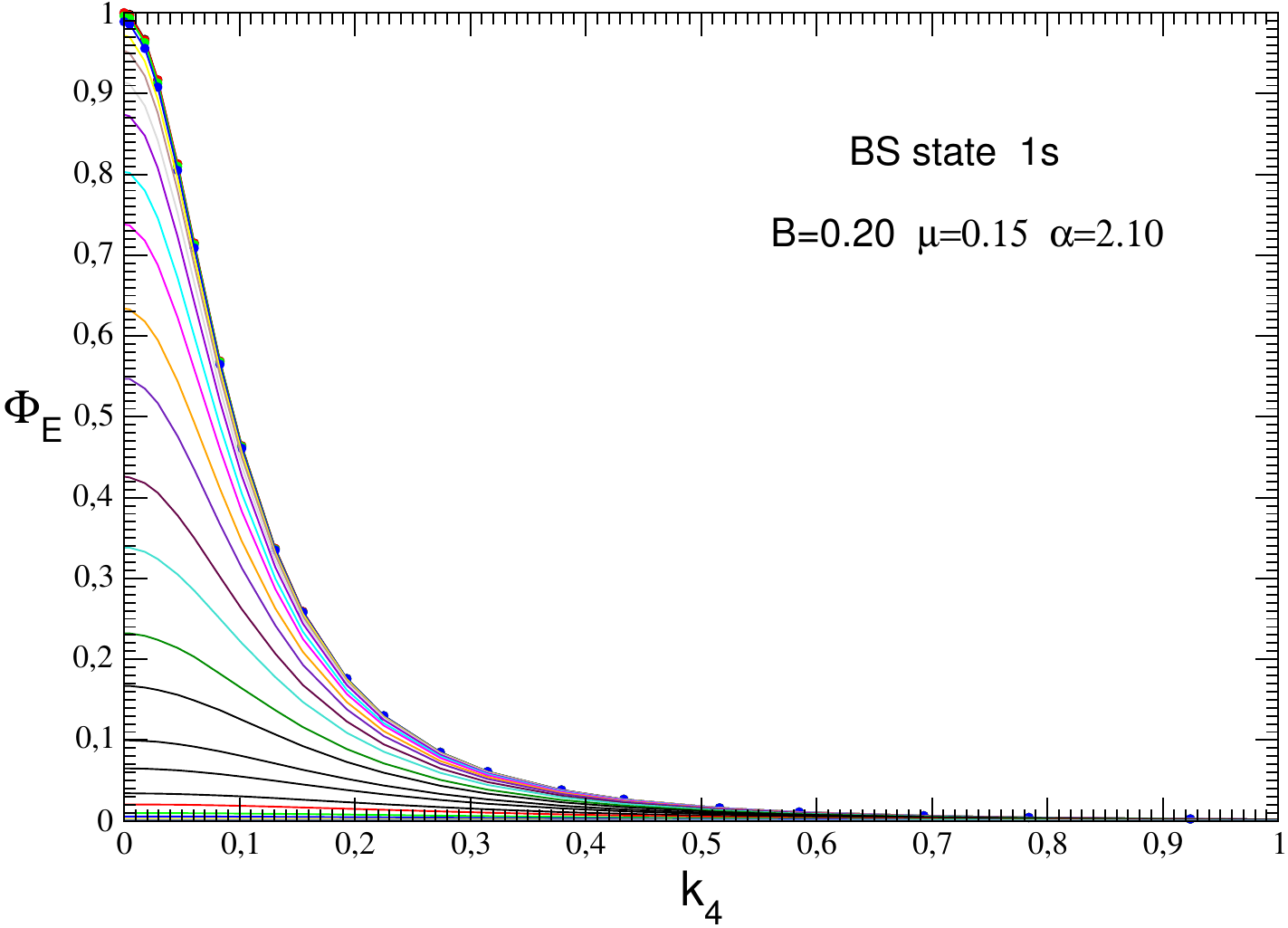}\hspace{0.2cm}
\includegraphics[width=6.cm]{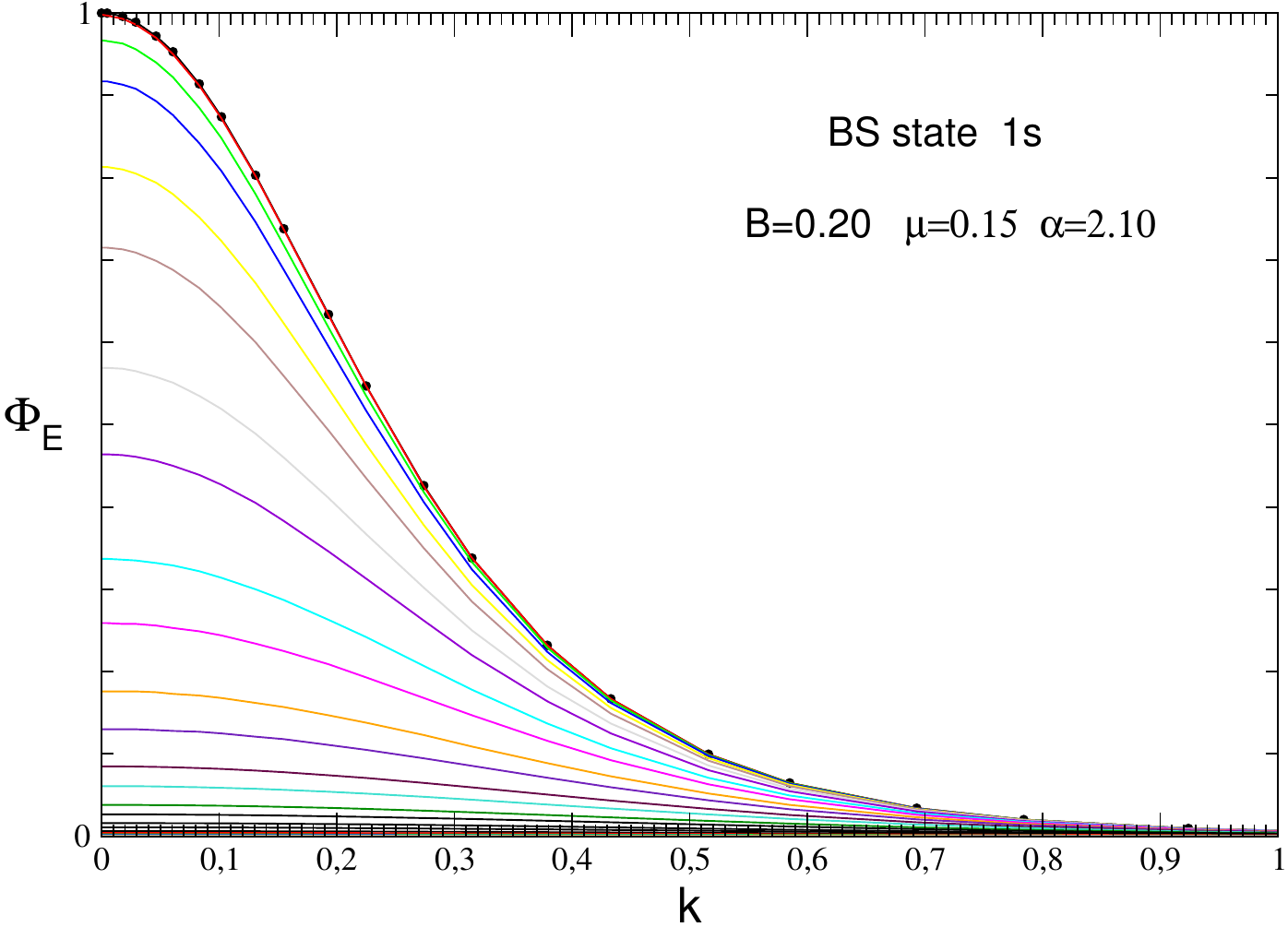}
\includegraphics[width=6.cm]{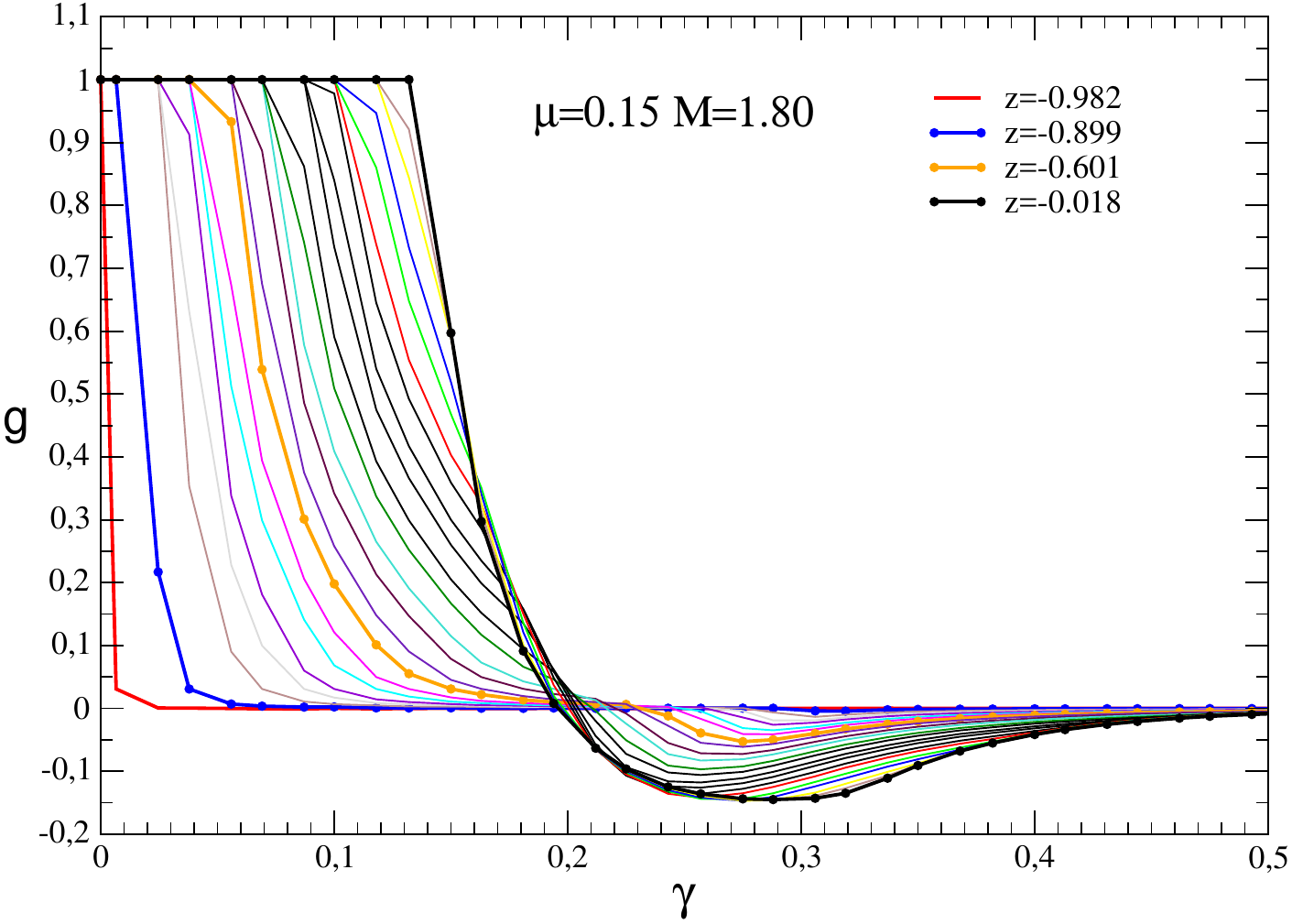}\hspace{0.2cm}
\includegraphics[width=6.cm]{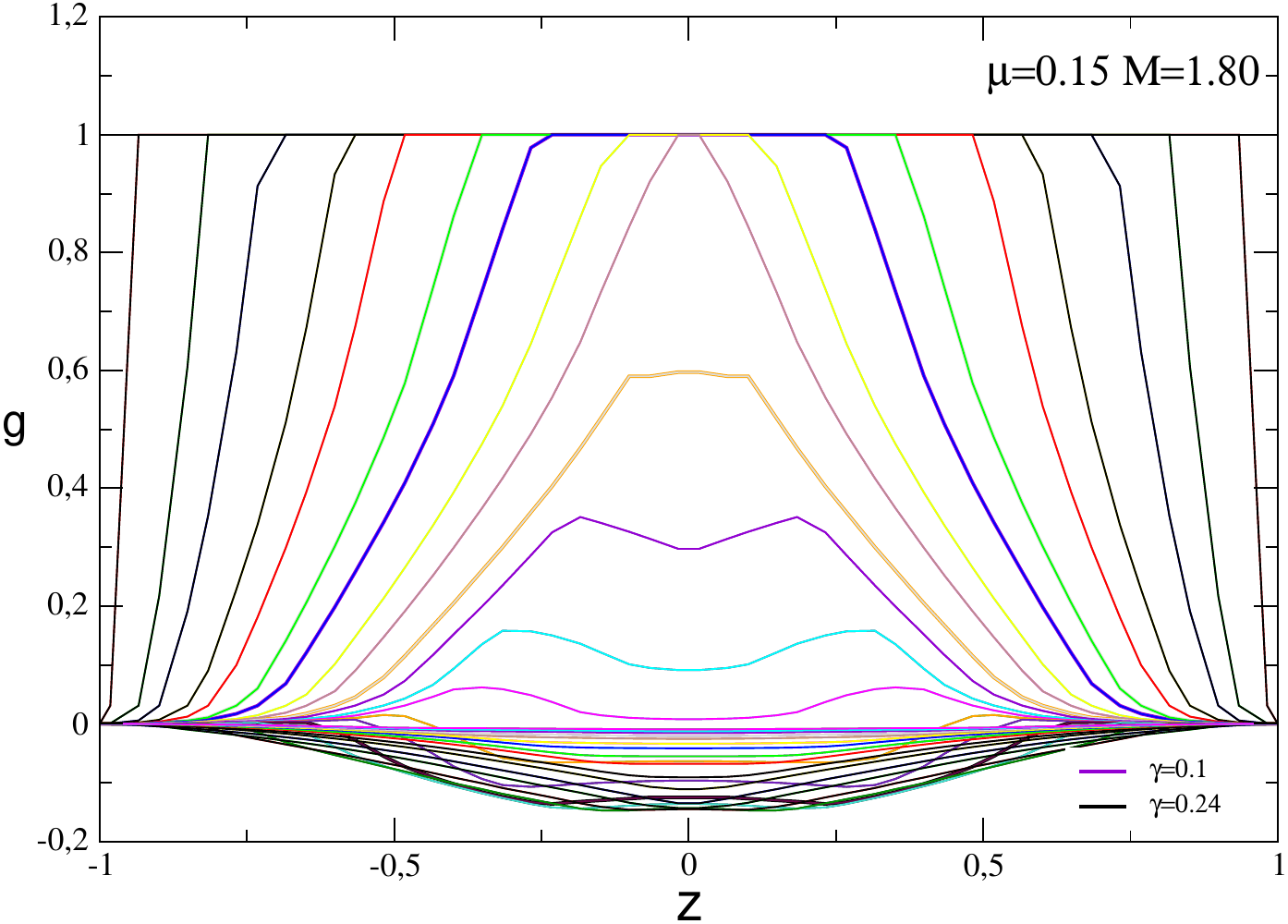}
\caption{Upper panel:  The Euclidean BS amplitude for $\mu=0.15$ and $B=0.2$. On the left,  as a function of $k_4$ 
for  selected values of $k$, and  on the right, as a function of $k$ for  selected values of $k_4$.
Lower panel: the same state  in terms of  the Nakanishi weight function $g(\gamma,z)$} \label{PhiE_g_mu_015_B_01}
\end{center}
\end{figure}

\begin{figure}[htbp]
\begin{center}
\includegraphics[width=4.cm]{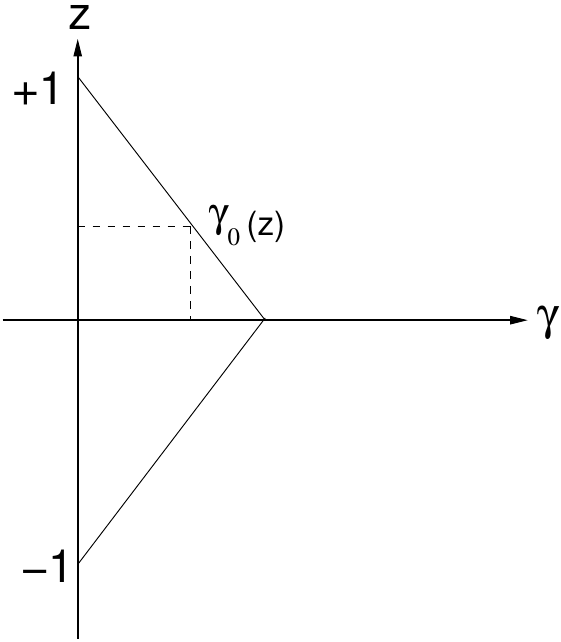}
\caption{Domain $\Delta$ of the ($\gamma,z$) where the solution $g(\gamma,z)$ is constant.}\label{Delta_z}
\end{center}
\end{figure}

We display in Fig. \ref{PhiE_g_mu_015_B_01} two representation of the same state, one (upper part) obtained with the BS solution in Euclidean space \eqref{EBSE}
and the other one (lower part) with the solution of eq. \eqref{gNg}. They correspond to $\mu=0.15$, a binding energy $B=0.2$, with the coupling constant  $\alpha=2.10$.
If the Euclidean solution in the upper part of the figure is a very smooth function of both arguments, 
the $\gamma$-dependence of the Nakanishi weight function $g$ is non-trivial. 
We can show analytically  \cite{CKKS_EPJA2} that $g$  is constant in  a triangular domain  $\Delta$ of the ($\gamma,z$) plane,   
 illustrated in Fig. \ref{Delta_z}, presents a cusp on its border and evolves continuously outside.
The analytic expression of this domain is  $\Delta= \{ (\gamma,z) \in R^2 : z\in[-1,+1] \;{\rm and} \;\gamma=\gamma_0(z) \}$ 
with $\gamma_0(z)$  given by
\begin{equation}\label{gamma0_z}
  \gamma_0(z)=(1-|z|){M^2\over 4m^2} \; \left(  {\mu^2\over m^2}+4{\mu\over M}\sqrt{1- {M^2\over 4m^2} } \sqrt{1-{\mu^2\over 4m^2} }   \right)
  \end{equation}

This peculiar  behaviour was  missed in our first publications \cite{bs1,bs2} because of some numerical instabilities
in the left-hand side kernel\footnote{We were solving at that time a generalized eigenvalue equation, formally writen as  $V_L\;g=V_R\; g$}
 and was only vaguely suggested in \cite{FSV_PRD89_2014} due to an unadapted basis set used in the numerical solution. 
It was however  well reproduced in  Figs. 2 and 3 of \cite{KW_PRD51_1995}.  It seems to be also  well reproduced in 
Fig. 5 of Ref. \cite{Jia_PRD109_2024}, although in this work the domain of constant $g$ is half a circle rather than a triangle,
maybe due to the used logarithmic scale or to some change of variable.  
Notice also that the negative part of $g$, visible in Fig. \ref{PhiE_g_mu_015_B_01} for $\gamma\sim0.2$, 
is totally absent in reference \cite{KW_PRD51_1995} and not clearly seen in \cite{Jia_PRD109_2024}.
On the contrary the results from  \cite{KSW_PRD56_1997}  are not understandable in terms of the previous analysis. 

When $\mu\ne0$ the  symmetry  of the 4D Coulomb problem is lost, as well as the $\kappa$ quantum number identifying the abnormal states
in the $\mu$=0 case.
The solutions of the BS equation are then labeled by a single quantum number $n$ which tells us nothing about
the normal or abnormal character of the state.
On another hand, the level ordering of the abnormal states in the $\mu$=0 case depends critically on the binding energy $B$ of the state.
As one can see in Fig. \ref{Fig_lambda_B}, for B=0.1, the first abnormal state ($\kappa$=2, $n$=1) (in green) is the 6th
excitation in the $\alpha$ spectrum, while for B=0.01 it  corresponds to the 11th... and for B=0 there is an infinity of normal states below the first abnormal one.
So, even for $\mu$=0, just by computing the $\alpha$ spectrum for a given $B$, there is no way to identify the abnormal state
 without study its wave function: there is an infinity of level crossing among
normal and abnormal states when $B$ is changed.

In a recent work we have extensively studied the  survival of the abnormal states in the massive case.  
The method and the results will be detailed in a forthcoming publication \cite{CKKS_EPJA2}.
We present in what follows a summary  of the main results concerning the very existence of such states.

The procedure is based on tracing the trajectories of a well-identified abnormal state for $\mu$=0 as a function of $\mu$,  
and on determining in this way  the ensemble of parameters ($\alpha,\mu$) allowing the existence of the abnormal states,
as well as their binding energies $B(\alpha,\mu)$. 
We will restrict here to the non-tachyonic domain  of the coupling constant: $1/4<\lambda={\alpha\over\pi} < 2$.

\begin{figure}[h]  
\begin{center}
 \includegraphics[width=6.cm]{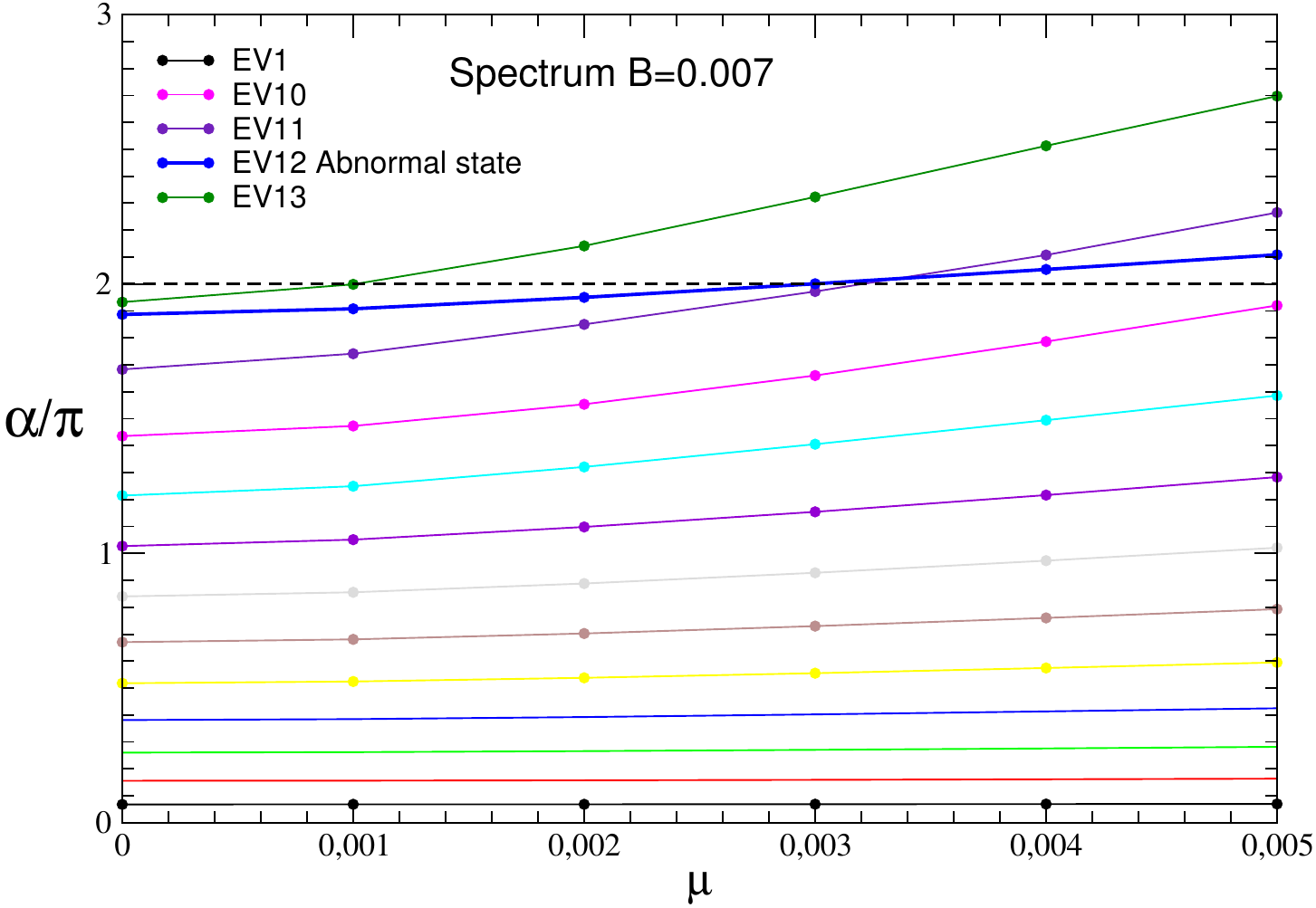}
\hspace{0.2cm} 
\includegraphics[width=6.5cm]{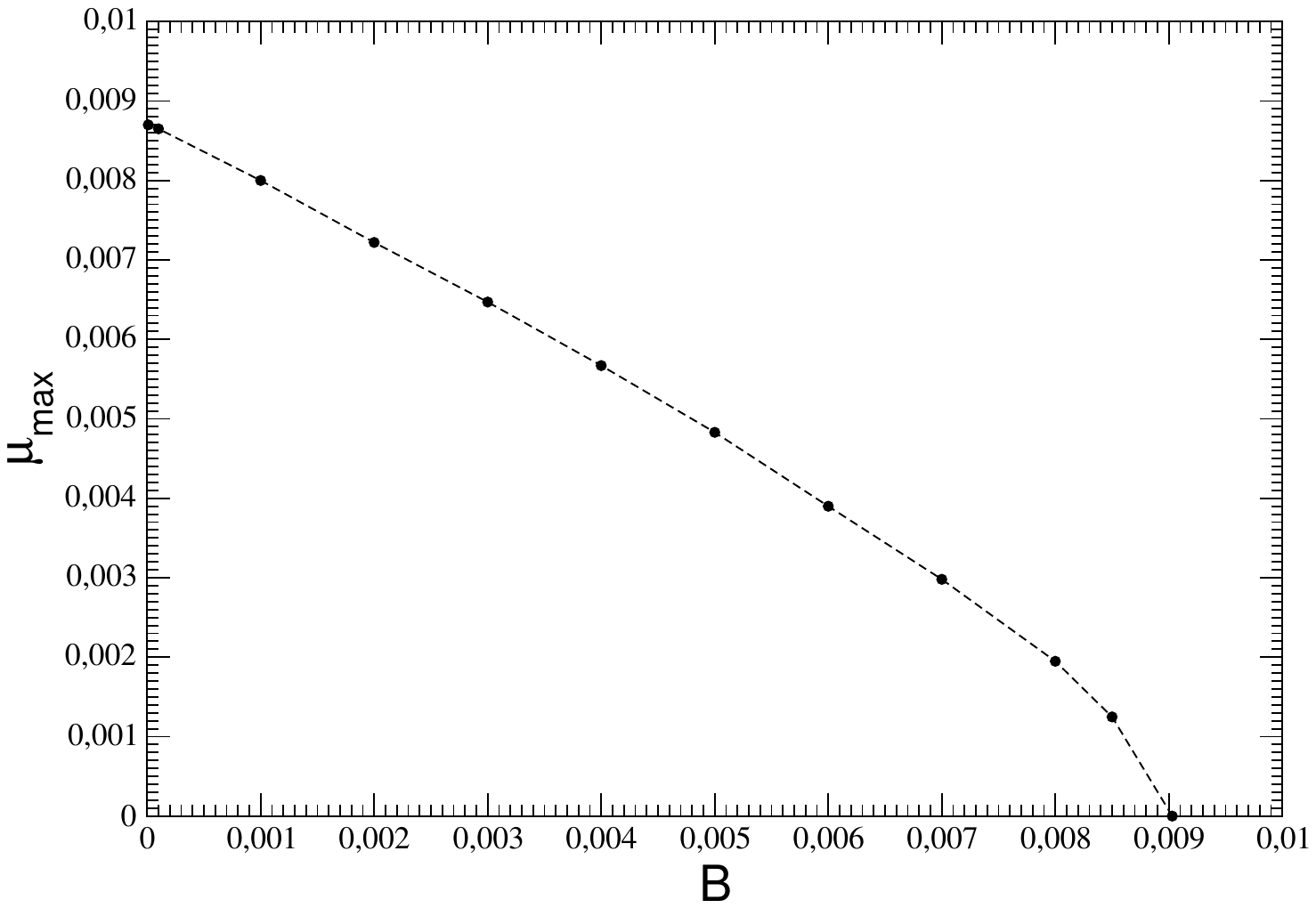}
\end{center} 
 \caption{Left panel: dependence of the $\alpha$-spectrum  as a function  of $\mu$, for   B=0.007. The horizontal dashed line  corresponds
  $\alpha_{max}(\mu)\approx2\pi$.
The trajectory   $\alpha(\mu)$ -- corresponding to the ground abnormal state (in blue solid line)  -- remains smaller than $\alpha_{max}(\mu)$  until $\mu\approx$0.0030.
The ground abnormal state, which at $\mu$=0 is the 12th excitation of $\alpha$ spectrum, displays a level crossing at $\mu\approx$0.0034 and becomes the 11th excited state.
Right panel: by repeating the study for several values of B one determines the domain $\mu_{max}(B)$.}\label{FIG_lambda_mu_B_0.007}        
  \end{figure}

An illustrative example is given in the left panel of Fig. \ref{FIG_lambda_mu_B_0.007}
for the ground abnormal state ($n$=1, $\kappa$=2) with binding energy $B$=0.007.
This state (in blue solid line) corresponds to the 12th excitation at $\mu$=0.
When $\mu$ is increased, the corresponding value of $\alpha$ increases until
it reaches the maximum allowed value  of the coupling constant $\alpha/\pi=2$. 
This determines,  the maximum allowed  exchanged mass $\mu$ for this state: $\mu_{max}\approx$0.0030.
{\bf This result constitutes the first  evidence, a  numerical proof of existence, of abnormal states in the W-C model with non-zero $\mu$}.
Such a possibility
was considered in Ref.  \cite{Naito_PThP40_1968}, in relation with its eventual contribution to the S-matrix.

By repeating this study for several values of $B$, one can determine $\mu_{max}(B)$, that is
the maximum value of $\mu$ compatible with a non-tachyonic ground state solution ($M^2>0$)  as a function of its binding energy $B$.
It is worth noticing here that for $\mu>0$ the non-tachyonic condition is not exactly given by $\alpha=2\pi$ but by a
a slightly larger value $\alpha_{max}(\mu)> 2\pi$ due to the short-range character of the $\mu>0$ interaction. 
Since the involved values of $\mu$ are very small, one can take for practical purposes $\alpha_{max}(\mu)= 2\pi$.
The $\mu_{max}(B)$ dependence is the essential ingredient in our study  and it is displayed in the right panel of  Fig. \ref{FIG_lambda_mu_B_0.007}.
The maximum allowed value of the exchanged mass is reached for $B=0$ and is $\mu_{max}(0)=0.087$ (in constituent units).

\begin{figure*}[htbp] 
\begin{center}
\includegraphics[width=12.5cm]{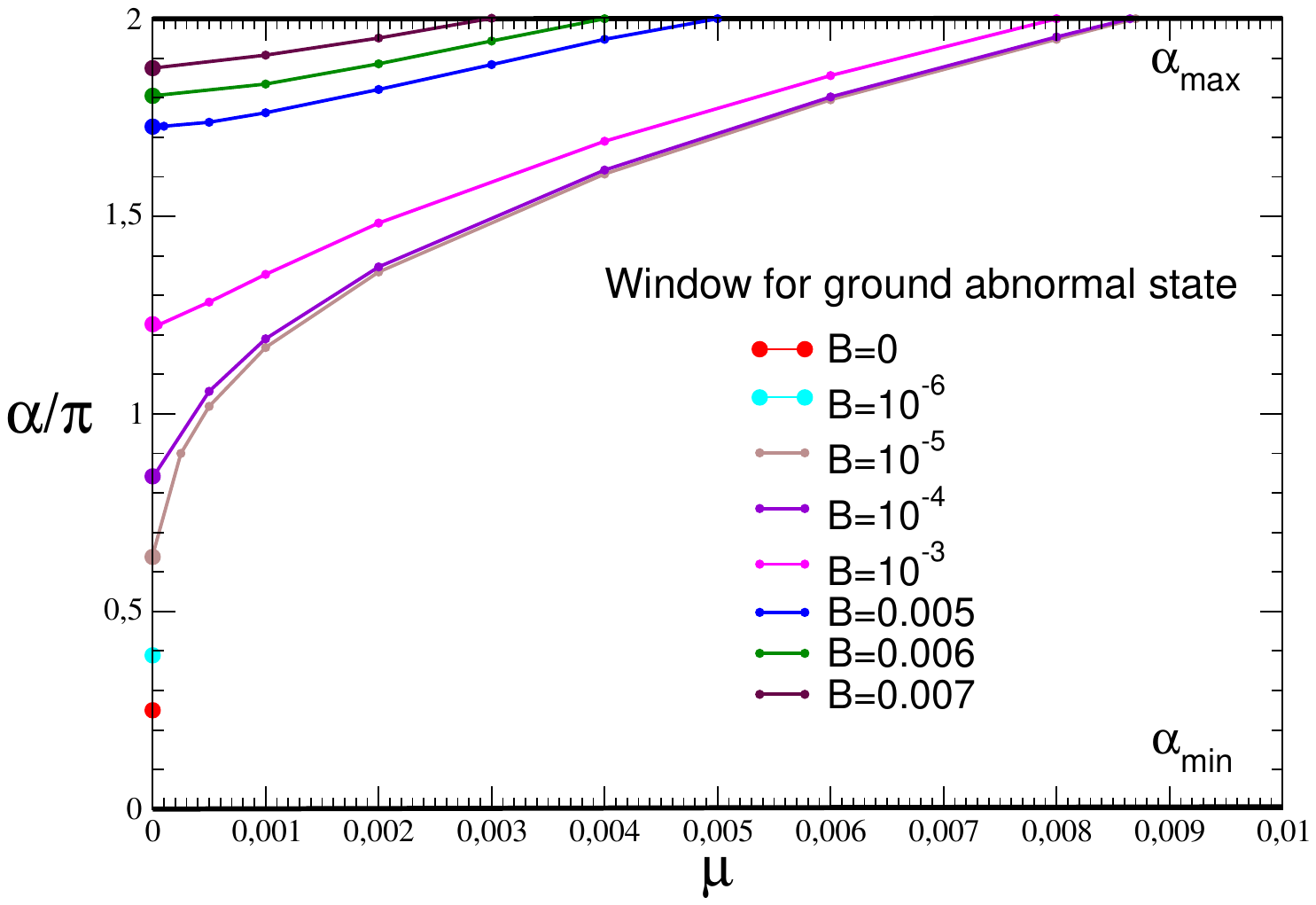}
  \end{center} 
 \caption{The parameter window ($\alpha,\mu$) of the ground abnormal state,  for several values of B. 
 For $B<10^{-5}$ the ground abnormal state is too high in the excitation spectrum to be numerically accessible:
 we give only the values of the coupling constant  corresponding to $\mu=0$.}\label{mumax_B_window}
\end{figure*}

We have repeated this study for several values of the binding energy and obtained the results of Fig. \ref{mumax_B_window}.
Our technology does not allow us to go below B=10$^{-5}$, essentially due to the difficulty of accurately computing higher
excited states. For $B$=10$^{-6}$ and $B$=0 we have just inserted the values for $\mu$=0.
On the other hand, it is worth noticing that the $\mu=0$ limit of the 
W-C model is non-analytic  and highly singular.
This is manifested already by the fact that the two-dimensional weigh function $g(\gamma,z)$ generates in this limit 
a $\delta(\gamma)$ function and there remains only the $z$-dependence. When computing the solutions for very small values
of $\mu$ we are faced to this embarrassing vicinity.
Let us also mention that the slope at $\mu=0$ of the $\alpha(\mu,B)$ represented at B=0 is infinite.

The curves displayed in  Fig. \ref{mumax_B_window}  delineate the parameter domain of the W-C model,  for which the abnormal states  exist.
It constitutes the main result of our work.
They allow us to draw two  conclusions.  The first one is that abnormal states exist for $\mu>0$. 
The second one is that, due to stability reasons of the theory, their binding energy is smaller than $B/m\approx0.00903$ and the exchanged mass is limited to very small values of $\mu$, $\mu/m<0.0087$.

\section{Conclusion}\label{Sec_Conclusions}

We  have reviewed  the main properties  of the ``abnormal solutions'' of the  Bethe--Salpeter equation with the Wick--Cutkosky model,
i.e., scalar particles with mass $m$  interacting via massless ($\mu$=0) exchange, as wells as its extension to the massive-exchange case.

These are low-energy ($B/m<0.009$)  solutions that exist in this  particular relativistic approach, but are absent in the non-relativistic limit (Coulomb problem).
Their position in the full spectrum of the model is totally decoupled from the normal solutions, which tend
to the standard Coulomb states, and require, even in the zero binding limit, a minimal value of the coupling constant, $\alpha_{min}=\pi/4$, to exist.
From this point of view the abnormal solutions behave as if they were created by an effective "massive photon".
In the  $\mu=0$ case, this is equivalent to a "massive photon" with an effective mass of $\mu_{eff}/m\approx 0.47$.
For the $\mu>0$ case, the value of $\mu_{eff}$ is roughly the same than for the massless case since the 
values of $\alpha_{min}$ are  practically unchanged with $\mu$ (See left panel of Fig \ref{FIG_lambda_mu_B_0.007}).

Discovered by Cutkosky \cite{Cutkosky_PR96_1954} soon after the formulation of the Bethe--Salpeter equation and its first solution
in Euclidean space by Wick \cite{Wick_PR96_1954},
we have given them \cite{Abnormal_1_2021} an intrinsic characterization  in terms of the small two-body norm 
of their valence wave function, which vanishes in the $B\to0$ limit.
This confers to them a genuine many body status.

We have presented new results concerning the massive-exchange case, where we have obtained the ensemble of parameters of the model, in particular
the values of the exchanged mass $\mu$, 
that allow the existence of such peculiar solutions without spoiling the model by tachyonic states ($M^2<0$).

As our previous  analysis shows, the reason for the existence of  abnormal states, 
dominated by multi-photon exchange, is the strong electrical field between constituents and therefore it can
be hardly affected by the eventual spin degrees of freedom which were not included in our consideration. 
The experimental creation and observation of these systems do not seem to be an easy task, but they would be of great interest.

\bigskip
\bmhead{Acknowledgments}
J.C. thanks the  financial support from FAPESP (Funda\c c\~ao de Amparo \`a Pesquisa do Estado de S\~ao Paulo) grant  2022/10580-3.  
H.S. acknowledges financial support from the EU research and innovation programme Horizon 2020,
under Grant agree\-ment No. 824093.



\end{document}